\documentclass[aps,prl,twocolumn,superscriptaddress,showpacs]{revtex4}
\usepackage{graphicx}% Include figure files
\usepackage{epstopdf,epsfig}
\usepackage{amsmath}
\usepackage{amsfonts}
\newcommand{\me}{\mathrm{e}}
\usepackage{times}

\begin{document}

\title{Quantum Hashing with the Icosahedral Group}

\author{Michele Burrello}
\affiliation{International School for Advanced Studies (SISSA),
  Via Beirut 2-4, I-34014 Trieste, Italy}
\affiliation{Istituto Nazionale di Fisica Nucleare, Sezione di Trieste, Italy}

\author{Haitan Xu}
\affiliation{Zhejiang Institute of Modern Physics, Zhejiang
University, Hangzhou 310027, P.R. China}

\author{Giuseppe Mussardo}
\affiliation{International School for Advanced Studies (SISSA),
  Via Beirut 2-4, I-34014 Trieste, Italy}
\affiliation{International Centre for Theoretical Physics (ICTP),
  I-34014 Trieste, Italy}
\affiliation{Istituto Nazionale di Fisica Nucleare, Sezione di Trieste, Italy}

\author{Xin Wan}
\affiliation{Asia Pacific Center for Theoretical Physics (APCTP),
Pohang, Gyeongbuk 790-784, Korea}
\affiliation{Department of Physics, Pohang University of Science and
Technology, Pohang, Gyeongbuk 790-784, Korea}
\affiliation{Zhejiang Institute of Modern Physics, Zhejiang
University, Hangzhou 310027, P.R. China}

\date{\today}

\begin{abstract}

  We study an efficient algorithm to hash any single qubit gate (or
  unitary matrix) into a braid of Fibonacci anyons represented by a
  product of icosahedral group elements. By representing the group
  elements by braid segments of different lengths, we introduce a
  series of {\em pseudo-groups}. Joining these braid segments in a
  renormalization group fashion, we obtain a Gaussian unitary ensemble
  of random-matrix representations of braids. With braids of length
  $O(\log^2(1/\varepsilon))$, we can approximate all SU(2) matrices to
  an {\it average} error $\varepsilon$ with a cost of
  $O(\log(1/\varepsilon))$ in time. The algorithm is applicable to
  generic quantum compiling.
 
\end{abstract}

\maketitle

Quantum gates are the building blocks for quantum circuits. A reliable
implementation of quantum computation would need a universal set of
fault-tolerant gates. How to use the set of universal gates to
construct quantum circuits is an important
question~\cite{nielsen}. The question also arises if we want to
simulate the circuits of the universal set by using those of another
set. The Solovay-Kitaev algorithm~\cite{kitaev} guarantees good
approximations to any desired gates, provided that a dense enough
$\epsilon$-net exists.  Instead of using quantum error-correction
codes, topological quantum
computation~\cite{kitaev03,freedman02,nayak08,brennen08,preskill}
proposes to realize fault-tolerant quantum gates by topology embedded
in hardware. In two-dimensional topological states of matter, a
collection of non-Abelian anyonic excitations with fixed positions
spans a multi-dimensional Hilbert space and, in such a space, the
quantum evolution of the multi-component wave function of the anyons
is realized by their braidings. The evolution can be represented by
non-trivial unitary matrices that implement quantum computation.  A
prototype of non-Abelian anyons is known as the Fibonacci anyons,
which exist in the Read-Rezayi quantum Hall state at filling fraction
$\nu = 3/5$~\cite{read99} (whose particle-hole conjugate is a
candidate for the observed $\nu = 12/5$ quantum Hall
plateau~\cite{xia04}) and in the non-Abelian spin-singlet state at
$\nu = 4/7$~\cite{ardonne99}.  In topological quantum computation, the
topology of the quantum braids precludes errors induced by local
noises; unfortunately, this does not eliminate the errors in
approximating quantum gates by braids.

\begin{figure}
  \centering
  \includegraphics[width=8cm]{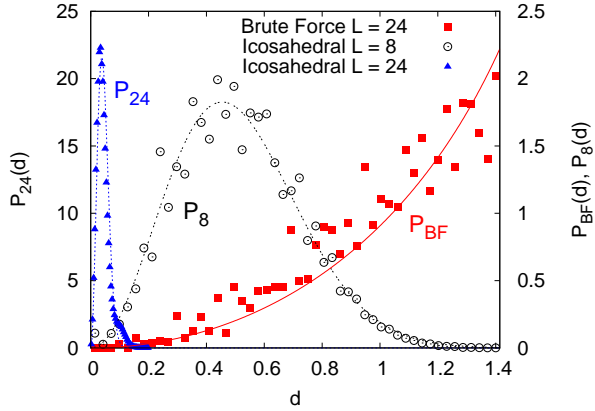}
  \caption{(color online) Probability distribution of distance
    $d$~\cite{distance} to the targeted identity matrix in the set of
    nontrivial braids that one samples in different
    algorithms. $P_{BF} (d)$ of the brute-force search (red solid
    squares) roughly follows $(4 / \pi) d^2 / \sqrt{1 - (d/2)^2}$,
    reflecting the three-sphere nature of the unitary matrix space
    (three independent parameters apart from an unimportant phase). In
    the pseudo icosahedral group approach ($n = 4$), distributions
    for $L = 8$ ($P_8$, black empty circles) and $L = 24$ ($P_{24}$,
    blue solid trangles) agree very well with the energy-level-spacing
    distribution of the unitary Wigner-Dyson ensemble of random
    matrices, $P_L(d) = (32 / \pi^2) (d^2/d_L^3) \me^{-(4/\pi)
      (d/d_L)^2}$.  $P_L(d)$ differ only by their corresponding
    average $d_L$ (not a fitting parameter), which decays
    exponentially as $L$ increases. Note $P_{24}(d)$ is roughly
    ten-times sharper and narrower than $P_8(d)$. }
  \label{fig:hist}
\end{figure}

Bonesteel {\it et al.} pioneered the implementation of quantum gates
using Fibonacci anyons with a brute-force search
algorithm~\cite{bonesteel05,simon06}, which finds the best
approximation to a unitary matrix $T$ in the set of all braids up to a
certain length $L$. As for all quantum computation schemes, the
complexity (thus inefficiency) in brute-force search is dictated by
the necessity to sample the whole space of unitary matrices with
almost equal weight, while the target gate is just a zero-measure
point inside. Thus the distance~\cite{distance} of the approximation
depends on $L$ as $e^{-L/\xi}$ (with $\xi \simeq
7.3$~\cite{xu08}). However, the run time grows exponentially in $L$,
rendering the algorithm impractical to achieve a distance below a
certain threshold.  In fact, the most probable braids generated by the
brute-force algorithm have the largest distance to the desired gate
due to the geometry of the unitary matrix space~\cite{distance}, as
illustrated in Fig.~\ref{fig:hist}. Subsequent
algorithms~\cite{xu08,xu082} enhance the sampling of the target point
by mapping it to a higher-dimensional object, although the search
remains time-consuming.  The inefficiency in these algorithms is also
reflected in the fact that a new unitary matrix needs a new
brute-force search, which is exponentially hard. The existing
implementation of the Solovay-Kitaev algorithm~\cite{hormozi07} is not
efficient enough in terms of either braid length or searching time.

The question is thus the following: can one implement a more efficient
search algorithm to find braids for single-qubit gates?  Technically,
we can think of a braid as an index to the corresponding unitary
matrix, which can be regarded as a definition, like in a
dictionary. Given an index, it is straightforward to find its
definition, but finding the index for a definition is exponentially
hard. In computer science, the task of quickly locating a data record
given its content (or search key) can be achieved by the introduction
of hash functions. In the context of topological quantum computation,
we thus name this task {\it topological quantum hashing}. In general,
such a hashing function, being imperfect, still maps a unitary matrix
to a number of braids rather than one.  But narrowing the search down
to only a fixed (rather than exponentially large) number of braids is
already a great achievement.

In this Letter, we explore topological quantum hashing with the finite
icosahedral group ${\mathcal I}$ and its algebra. The building blocks
of the algorithm are a {\em preprocessor} and a {\em main processor}:
the aim of the preprocessor is to give an initial approximation
$\tilde T$ of the target gate $T$, while that of the main processor is
to reduce the discrepancy between $T$ and $\tilde T$ with extremely
high efficiency. We discuss the iteration of the algorithm in a
renormalization group fashion and the results which follow from 
this approach. The algorithm is also applicable to generic quantum compiling and, remarkably, its efficiency can be quantified 
using random matrix theory. 

We illustrate our algorithm with Fibonacci anyons (denoted as $\phi$, with a fusion
rule $\phi \times \phi = 1 + \phi$, where 1 is the
vacuum)~\cite{bonesteel05,simon06,xu08,xu082,hormozi07}. If we create
two pairs of $\phi$ (illustrated graphically by dots) out of the
vacuum, both pairs (small ellipses) must have the same fusion outcome,
1 or $\phi$, forming a qubit (large ellipse), in which the braiding of
$\phi$'s can be generated by two fundamental braiding matrices

\begin{figure}[h!]
\begin{minipage}[c]{2cm}
\centering
\includegraphics[height=1.2cm]{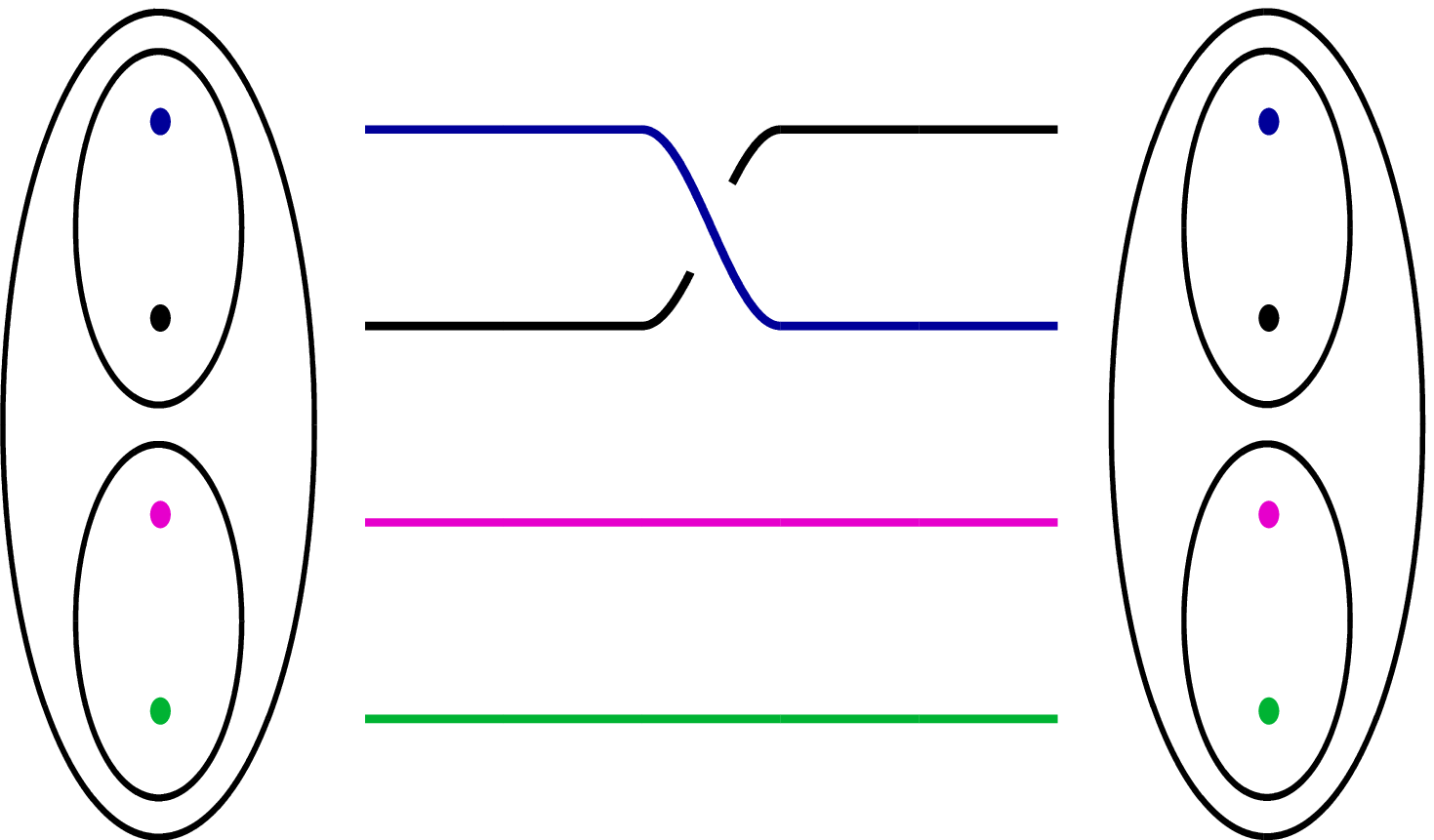}
\end{minipage}
\begin{minipage}[c]{6cm}
\vspace{-0.2cm}
\begin{equation}
\label{eq:s2}
\sigma_1 = \left [ 
\begin{array}{cc}
\me^{-i 4 \pi / 5} & 0 \\
0 & -\me^{-i 2 \pi / 5}
\end{array}
\right ], 
\end{equation}
\end{minipage}
\end{figure}
\vspace{-0.5cm}
\begin{figure}[h!]
\begin{minipage}[c]{2cm}
\centering
\includegraphics[height=1.2cm]{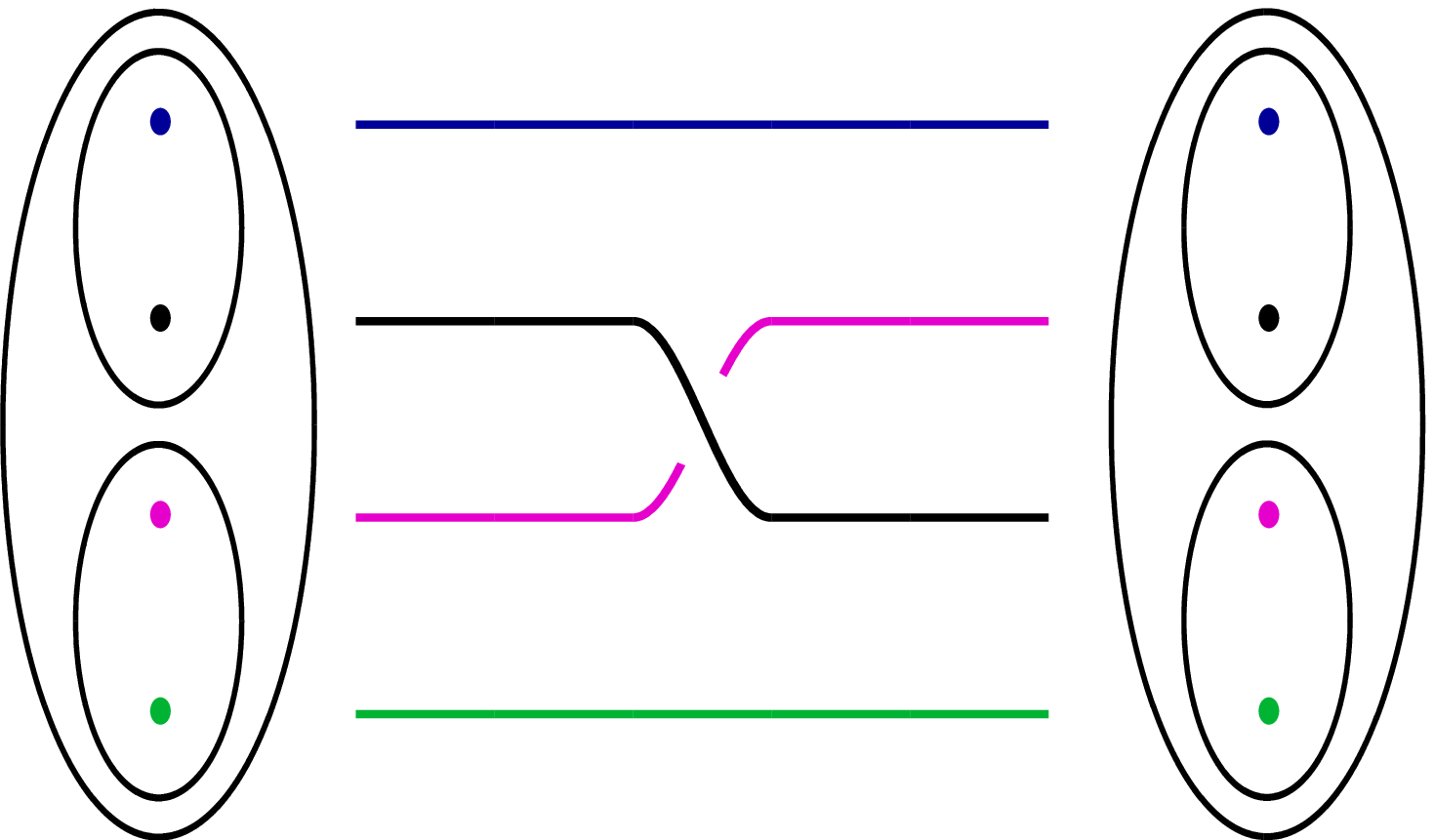}
\end{minipage}
\begin{minipage}[c]{6cm}
\vspace{-0.2cm}
\begin{equation}
\sigma_2 = \left [ 
\begin{array}{cc}
-\tau \me^{-i \pi / 5} & -\sqrt{\tau} \me^{i 2 \pi / 5} \\
 -\sqrt{\tau} \me^{i 2 \pi / 5} & -\tau
\end{array}
\right ], 
\end{equation}
\end{minipage}
\end{figure}

\noindent
and their inverses $\sigma_1^{-1}$, $\sigma_2^{-1}$. Here $\tau =
(\sqrt{5} - 1)/2$. The matrix representation generates a four-strend
braid group $B_4$ (or an equivalent three-strand braid group $B_3$):
this is an infinite dimensional group consisting of all possible
sequences of length $L$ of the above generators and with increasing
$L$ the whole set of braidings generates a dense cover of the SU(2)
single-qubit rotations. Earlier
works~\cite{bonesteel05,xu08,xu082,hormozi07} have demonstrated that
the two-qubit gate construction can be mapped to the single-qubit gate
construction; thus, we will not discuss the construction of two-qubit
gates here.

{\em Icosahedral group}. The icosahedral rotation group ${\mathcal I}$
of order 60 is the largest finite subgroup of SU(2) excluding
reflection. Therefore, it has been often used to replace the full
SU(2) group for practical purposes, as for example in earlier Monte
Carlo studies of SU(2) lattice gauge theories~\cite{rebbi80}, and
this motivated us to apply the icosahedral group representation in the
braid construction. ${\mathcal I}$ is composed by the 60 rotations
around the axes of symmetry of the icosahedron (platonic solid with
twenty triangular faces) or of its dual polyhedron, the dodecahedron
(regular solid with twelve pentagonal faces); there are six axes of
the fifth order, ten of the third and fifteen of the second. Let us
for convenience write ${\mathcal I} = \{g_0, g_1, ...,g_{59}\}$, where
$g_0 = e$ is the identity element.

\begin{figure}
  \centering
\begin{minipage}[c]{8cm}
\centering
\includegraphics[width=6cm]{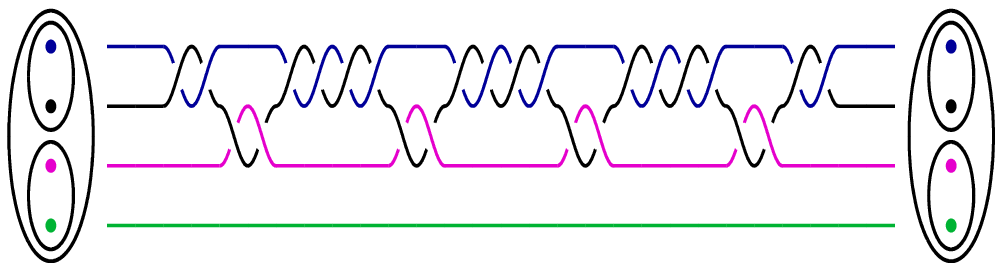}
\vspace{-0.35cm}
\begin{displaymath}
\label{eq:seq24}
\sigma_1^{-2} \sigma_2^2 \sigma_1^{-4} \sigma_2^2 \sigma_1^{-4} \sigma_2^2 \sigma_1^{-4} \sigma_2^2 \sigma_1^{-2}
\end{displaymath}
\end{minipage}
\caption{(color online) Approximation to the $-iX$ gate (an element of
  the icosahedral group) in terms of braids of the Fibonacci anyons of
  length $L = 24$ in the graphic representation. In this example the
  error is 0.0031.}
\label{examplebraiding}
\end{figure}

\begin{figure*}
  \centering
  \includegraphics[width=16cm]{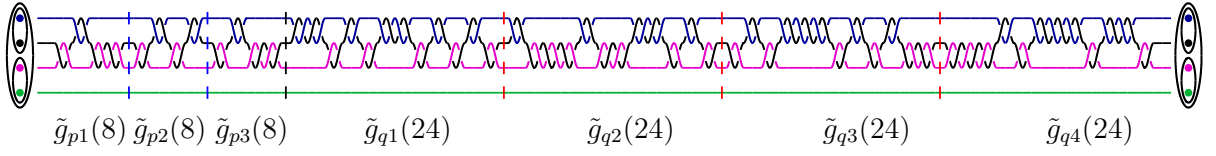}
  \caption{(color online) The graphic representation of the braid
    approximating the target gate $iZ$ in the icosahedral group
    approach with a preprocessor of $l = 8$ and $m = 3$ and a main
    processor of $L = 24$ and $n = 3$. To emphasize the structure, we
    skip the explicit braid sequence but mark the segments only, among
    which $\tilde{g}_{p1} \tilde{g}_{p2} \tilde{g}_{p3} \approx iZ$
    and $\tilde{g}_{q1} \tilde{g}_{q2} \tilde{g}_{q3} \approx
    \tilde{g}_{q4}^{-1}$ up to a phase. The braid (with a reduced
    length of 98 due to accidental cancellations where the component
    braids connect) has an error of 0.00099~\cite{sourcecode}. }
    \label{fig:braid120}
\end{figure*}

Thanks to the homomorphism between SU(2) and $SO(3)$, we start by
associating a $2 \times 2$ unitary matrix to each group element. In
other words, each group element can be approximated by a braid of
Fibonacci anyons of a certain length $N$ using the brute-force
search~\cite{bonesteel05} and neglecting an overall phase. In this
way, we obtain an approximate representation in SU(2) of the
icosahedral group, $\tilde{\mathcal{I}}(N) = \{ \tilde{g}_0(N),
\tilde{g}_1(N), \dots, \tilde{g}_{59}(N) \}$.  Choosing, for instance,
a fixed braid length of $N = 24$, the distance (or error) of each
braid representation to its corresponding exact matrix representation
varies from 0.003 to 0.094 (see Fig.~\ref{examplebraiding} for an
example).

We point out that the 60 elements of $\tilde {\mathcal I}(N)$ (for any
finite $N$) do not close any longer the composition laws of ${\mathcal
  I}$; in fact, they form a {\it pseudo-group}, not a group,
isomorphic to ${\mathcal I}$ only in the limit $N \rightarrow \infty$.
In other words, if the composition law $g_i g_j = g_k$ holds in the
original icosahedral group, the product of the corresponding elements
$\tilde{g}_i(N)$ and $\tilde{g}_j(N)$ is not $\tilde{g}_k(N)$,
although it can be very close to it for large enough
$N$. Interestingly, the distance between the product $\tilde{g}_i(N)
\,\tilde{g}_j(N)$ and the corresponding element $g_k$ of ${\mathcal
  I}$ can be linked to the Wigner-Dyson distribution, which we will
discuss later.

Using the pseudo-group structure of $\tilde {\mathcal
  I}$, we can generate a set $\mathcal{S}$ made of a large
number of braids only in the vicinity of the identity matrix: this is
a simple consequence of the original icosahedral group algebra, in
which the composition laws allow us to obtain the identity group
element in various ways. The set $\mathcal{S}$ is instrumental to
achieve an important goal, i.e. to search among the elements of $S$
the best correction to apply to a first rough approximation of the
target single-qubit gate $T$ we want to hash.  We can create such a
set, labeled by $\mathcal{S}(L,n)$, considering all the possible
ordered products $\tilde{g}_{i_1}(L) \tilde{g}_{i_2}(L) \dots
\tilde{g}_{i_n}(L)$ of $n\ge 2$ elements of $\tilde{\mathcal{I}}(L)$
of length $L$ and multiplying them by the matrix
$\tilde{g}_{i_{n+1}}(L) \in \tilde{\mathcal{I}}(L)$ such that
$g_{i_{n+1}}=g_{i_n}^{-1}\dots g_{i_2}^{-1}g_{i_1}^{-1}$. In this way
we generate all the possible combinations of $n+1$ elements of
${\mathcal I}$ whose result is the identity, but, thanks to the errors
that characterize the braid representation $\tilde{\mathcal{I}}$, we
obtain $60^n$ small rotations in SU(2), corresponding to braids of
length $(n+1)L$.

{\em The hashing procedure}. The first step in the hashing procedure
of the target gate is to find a rough braid representation of $T$
using a preprocessor, which associates to $T$ the element in
$[\tilde{\mathcal{I}}\left(l\right)]^m$ (of length $m \times l$) that
best approximates it. Thus we obtain a starting braid
\begin{equation}
  \tilde T^{l,m}_0  = \tilde{g}_{j_1}\left(l\right) \tilde{g}_{j_2}\left(l\right) \dots
  \tilde{g}_{j_m}\left(l\right)
\end{equation}
characterized by an initial error we want to reduce.  The preprocessor
procedure relies on the fact that choosing a small $l$ we obtain a
substantial discrepancy between the elements $g$ of the icosahedral
group and their representatives $\tilde g$. Due to these random errors
the set $[\tilde{\mathcal{I}}(l)]^m$ of all the products $\tilde{g}_{j_1}
\tilde{g}_{j_2} \dots \tilde{g}_{j_m}$ is well spread all over SU(2)
and can be considered as a random discretization of this group.

\begin{figure}
  \centering
  \includegraphics[width=8cm]{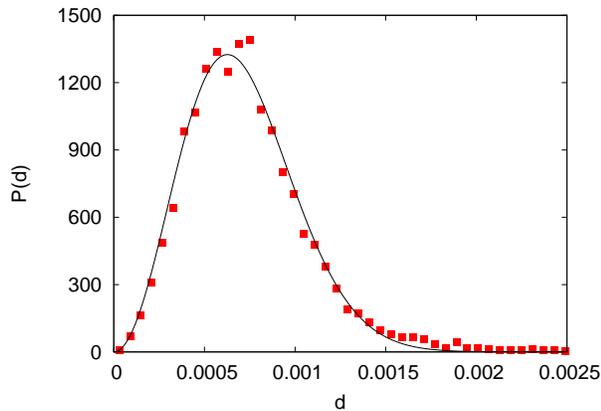}
  \caption{Probability distribution of $d$ in 10,000 random tests
    using the icosahedral group approach with a preprocessor of $l =
    8$ and $m = 3$ and a main processor of $L = 24$ and $n = 3$. The
    total length of the braids (neglecting accidental cancellations
    when component braids connect) is 120. The trend agrees with the
    unitary Wigner-Dyson distribution (solid line) with an average
    error $7.1 \times 10^{-4}$. }
  \label{fig:error}
\end{figure}

In the main processor we use the set of fine rotations
$\mathcal{S}(L,n)$ to efficiently reduce the error in $\tilde
{T}^{l,m}_0$. Multiplying $\tilde{T}^{l,m}_0$ by all the elements of
$\mathcal{S}(L,n)$, we generate $60^n$ possible braid representations
of $T$ :
\begin{equation} \label{approx}  
\tilde{T}^{l,m}_0 \tilde{g}_{i_1} \tilde{g}_{i_2} \dots
  \tilde{g}_{i_n} \tilde{g}_{i_{n+1}} 
\end{equation}
Among these braids of length $(n+1)L + ml$, we search the one which
minimizes the distance with the target gate $T$. This braid,
$\tilde{T}^{l,m}_{L,n}$, is the result of our
algorithm. Fig.~\ref{fig:error} shows the distribution of final errors
for 10,000 randomly selected target gates obtained with a preprocessor
of $l = 8$ and $m = 3$ and a main processor of $L = 24$ and $n = 3$.

To illustrate our algorithm, it is useful to consider a concrete
example: suppose we want to find the best braid representation of the
target gate
\begin{equation} \label{ex1}
T= iZ = \begin{pmatrix} i & 0 \\ 0 & -i \end{pmatrix}
\end{equation}
Out of all combinations in $[\tilde {\mathcal{I}}(8)]^3$, the
preprocessor selects a $\tilde T^{8,3}_0 = \tilde g_{p_1} (8) \tilde
g_{p_2}(8) \tilde g_{p_3}(8)$, which minimizes the distance to $T$ to
0.038. Applying now the main processor, the best rotation in
$\mathcal{S}(24,3)$ that corrects $\tilde T^{8,3}_0$ is given by a
$\tilde g_{q_1} (24) \tilde g_{q_2}(24) \tilde g_{q_3}(24) \tilde
g_{q_4}(24)$, where $ g_{q_4} = g_{q_3}^{-1} g_{q_2}^{-1}
g_{q_1}^{-1}$. The resulting braid~\cite{sourcecode} is then
represented by
\begin{multline} 
  \label{ex2} T^{8,3}_{24,3} = \tilde g_{p_1} (8) \tilde g_{p_2}(8) \tilde
  g_{p_3}(8) \tilde g_{q_1}(24) \tilde g_{q_2}(24) \tilde g_{q_3}(24) 
  \tilde g_{q_4}(24)  \\
= \begin{pmatrix} -0.0004 + 1.0000 i & -0.0007 - 0.0005 i \\ 
   0.0007 - 0.0005 i & -0.0004 - 1.0000 i
    \end{pmatrix} e^{\frac{4}{5}\pi i } 
\end{multline}
for the special set of $p$'s and $q$'s and, apart from an overall phase,
the final distance is reduced to 0.00099 (Fig.~\ref{fig:braid120}).

{\em Relationship with random matrix theory}. The distribution of the
distance between the identity and the so-obtained braids has an
intriguing connection to the Gaussian unitary ensemble of random
matrices, which helps us to understand how close we can approach the
identity in this way, i.e. the efficiency of the hashing
algorithm. Let us analyze the group property deviation for the
pseudo-group $\tilde{\mathcal{I}}(N)$ for braids of length $N$. One
can write $\tilde{g}_i = g_i \me^{i \Delta_i}$, where $\Delta_i$ is a
Hermitian matrix, indicating the small deviation of the finite braid
representation to the corresponding SU(2) representation for an
individual element.  For a product of $\tilde{g}_i$ that approximate
$g_i g_j \cdots g_{n+1}=e$, one has
\begin{equation}
\tilde{g}_i \tilde{g}_j  \cdots \tilde{g}_{n+1} = 
g_i \me^{i \Delta_i} g_j \me^{j \Delta_j}  
\cdots g_{n+1} \me^{i \Delta_{n+1}}
= \me^{i H_{n}},
\end{equation}
where $H_{n}$, related to the accumulated deviation, is 
\begin{multline}
\label{eq:rg}
H_{n} = g_i \Delta_i g_i^{-1} 
+ g_i g_j \Delta_j g_j^{-1} g_i^{-1} + \cdots  \\
+ g_i g_j \cdots g_n\Delta_n g_n^{-1} \cdots g_j^{-1}g_i^{-1} 
+\Delta_{n+1}+ O(\Delta^2).
\end{multline}
The natural conjecture is that, for a long enough sequence of matrix
product, the Hermitian matrix $H_n$ tends to a random matrix
corresponding to the Gaussian unitary ensemble. This is plausible as
$H_n$ is a Hermitian matrix that is the sum of random initial
deviation matrices with random unitary transformations. A direct
consequence is that the distribution of the eigenvalue spacing $s$
obeys the Wigner-Dyson form~\cite{mehta},
\begin{equation}
P(s) = {\frac{32}{\pi^2 s_0}} \left ({\frac{s}{s_0}}\right )^2 
\me^{-(4/\pi) (s/s_0)^2},
\end{equation}
where $s_0$ is the mean level spacing.  For small enough deviations,
the distance of $H_n$ to the identity, $d \left (1, \me^{iH_n} \right
) = \Vert H_n \Vert + O \left (\Vert H_n \Vert^3 \right )$, is
proportional to the eigenvalue spacing of $H$ and, therefore,
should obey the same Wigner-Dyson distribution. The conjecture above
is indeed well supported by our numerical analysis, even for $n$ as
small as 3 or 4 (see Fig.~\ref{fig:hist}). One can show that the final
error of $\tilde{T}^{l,m}_{L,n}$ also follows the Wigner-Dyson
distribution (as illustrated in Fig.~\ref{fig:error}) with an average
final distance $f \sim 60^{n/3}/\sqrt{n+1}$ times smaller than the
average error of $\tilde {T}^{l,m}_0$, where the factor 60 is given by
the order of the icosahedral group.  With a smaller finite subgroup of
SU(2), we would need a greater $n$ to achieve the same reduction.

{\em Conclusions}.  In this paper we have demonstrated that the
problem of compiling an arbitrary SU(2) qubit gate $T$ in terms of
Fibonacci anyons can be solved efficiently by using hashing functions
based on the 60 elements of the icosahedral group ${\mathcal I}$ and
their composition laws. Our procedure can be generalized to other
anyonic models, different quantum computational schemes, and in
principle to multi-qubit gates.

The hashing algorithm uses a light brute-force search up to $L = 24$
to initialize the 60 elements of $\mathcal{I}$ with an average
precision of about 0.02. The remaining search operations are based on
the composition laws of the group ${\mathcal I}$, which do not need 
any longer to exhaust the exponentially growing number of possibilities as $L$
increases. Indeed, it takes {\it less than a second} on a 3~GHz Intel E6850
processor to reach an {\it average} precision of $7.1 \times 10^{-4}$
(Fig.~\ref{fig:error}) for an arbitrary gate~\cite{sourcecode}.

We can further improve the precision with additional iterations in the
main processor, as we move exponentially down in error scales in a
renormalization group fashion. For that we need longer braid
representations of $\mathcal{I}$, which must be obtained
separately, e.g., by the brute-force search, and can be stored for all
future uses. It follows that $q$ iterations reduce the average error
by $f^q$ within a run time linear in $q$. To achieve an error smaller
than a given $\varepsilon$, one needs $q \sim \log(1/\varepsilon)$
consecutive iterations. Therefore, the run time grows as $T \sim \log
(1/\varepsilon)$, better than the poly-logarithmic time of the
efficient implementation of the Solovay-Kitaev
algorithm~\cite{dawson06}. The iterative hashing algorithm generates a
final braid of length $O(\log^2(1/\varepsilon))$, competing favorably
with the results of other efficient quantum compiling
algorithms~\cite{nielsen,dawson06}. We hope that the quantum hashing
algorithm, with potential improvements and hybridizations with other
algorithms, introduces a new direction for efficient quantum
compiling.

This work is supported the grants INSTANS (from ESF), 2007JHLPEZ (from
MIUR), and the PCSIRT Project No. IRT0754 (from MoE, PRC).
X.W. acknowledges the Max Planck Society and the Korea Ministry of
Education, Science and Technology for the support of the Independent
Junior Research Group at APCTP.

During the write-up of this Letter, we noticed a recent
paper~\cite{mosseri08} which discusses a geometrical approach with
binary polyhedral groups.


\begin{thebibliography}{99}

\bibitem{nielsen} M. A. Nielsen and I. L. Chuang, {\it Quantum
    Computation and Quantum Information}, Cambridge University Press
  (2000), Chapter 4. and Appendix 3. 

\bibitem{kitaev} A. Yu. Kitaev, A. H. Shen, and M. N. Vyalyi, {\it
    Classical and Quantum Computation}, Am. Math. Soc. (2002), Section
  8.

\bibitem{kitaev03} A. Yu. Kitaev, Ann. Phys. {\bf 303}, 2 (2003); {\it
    ibid.} {\bf 321}, 2 (2006).

\bibitem{freedman02} M. Freedman, M. Larsen, and Z. Wang,
  Commun. Math. Phys.  {\bf 227}, 605 (2002); {\bf 228}, 177
  (2002); M. Freedman, A. Kitaev, and Z. Wang, {\it ibid.} {\bf 227},
  587 (2002).

\bibitem{nayak08} C. Nayak {\it et al.}, Rev. Mod. Phys. {\bf 80},
1083 (2008).

\bibitem{brennen08}
G. K. Brennen and J. K. Pochos, Proc. R. Soc. A, {\bf
464}, 1 (2008).

\bibitem{preskill} J. Preskill, {\it Lecture Notes on
 Topological
  Quantum Computation}; available online at
  www.theory.caltech.edu/$\sim$preskill/ph219/topological.pdf.

\bibitem{read99}
N. Read and E. H. Rezayi, Phys. Rev. B {\bf 59}, 8084 (1999).

\bibitem{xia04}
J. S. Xia, {\it et al.}, Phys. Rev. Lett. {\bf 93}, 176809 (2004).

\bibitem{ardonne99} E. Ardonne and K. Schoutens, Phys. Rev. Lett. {\bf
  82}, 5096 (1999); E. Ardonne {\it et al.}, 
% N. Read, E. Rezayi, and K. Schoutens,
  Nucl. Phys. B {\bf 607}, 549 (2001); E. Ardonne and K. Schoutens,
  Ann. Phys. {\bf 322}, 201 (2007).

\bibitem{bonesteel05}
N. E. Bonesteel {\it et al.},
%N. E. Bonesteel, L. Hormozi, G. Zikos, and S. H. Simon, 
Phys. Rev. Lett. {\bf 95}, 140503 (2005).

\bibitem{simon06}
S. H. Simon {\it et al.},
Phys. Rev. Lett. {\bf 96}, 070503 (2006).

\bibitem{xu08}
H. Xu and X. Wan, Phys. Rev. A {\bf 78}, 042325 (2008).

\bibitem{distance} The distance $d$ (also referred to as error)
  between two gates (or their matrix representations) $U$ and $V$ is
  the operator norm distance $d\left( U,V\right) \equiv \Vert U - V
  \Vert = \sup _{\Vert \psi \Vert = 1} \Vert \left( U-V\right) \psi
  \Vert$. Note $\max(d) = \sqrt{2}$ when $U \psi$ is orthogonal to $V
  \psi$.

\bibitem{xu082}
H. Xu and X. Wan, Phys. Rev. A {\bf 80}, 012306 (2009).

\bibitem{hormozi07}
L. Hormozi {\it et al.},
%L. Hormozi, G. Zikos, N. E. Bonesteel, and S. H. Simon, 
Phys. Rev. B {\bf 75}, 165310 (2007).

\bibitem{rebbi80}
C. Rebbi, Phys. Rev. D {\bf 21}, 3350 (1980); 
%\bibitem{petcher80}
D. Petcher and D. H. Weingarten, Phys. Rev. D {\bf 22}, 2465 (1980);
%\bibitem{bhanot87}
G. Bhanot, K. Bitar, and R. Salvador, Phys. Lett. B {\bf 188}, 246 (1987).

\bibitem{sourcecode} Object-oriented source codes available for
  download at http://sites.google.com/site/braidanyons/.

\bibitem{mehta}
M. Mehta, {\it Random Matrices}, 2nd ed. (Academic, San Diego, 1991).

\bibitem{dawson06} C. M. Dawson and M. A. Nielsen,
  Quant. Info. Comp. {\bf 6}, 81 (2006).

\bibitem{mosseri08}
R. Mosseri, J. Phys. A: Math. Theor. {\bf 41}, 175302 (2008). 

\end{thebibliography}
\end{document}